\documentclass[conference]{IEEEtran}
\usepackage[table,xcdraw]{xcolor}
\usepackage{graphicx}
\usepackage{epstopdf}
\usepackage{amssymb}
\usepackage[T1]{fontenc}
\usepackage[font=small,labelfont=bf,tableposition=top]{caption}
\usepackage[numbers]{natbib}
\makeatletter
\begin{document}
\title{Analyzing the activity of a person in a chat by combining network analysis and fuzzy logic}
\author{\IEEEauthorblockN{Sude Tavassoli}
\IEEEauthorblockA{Graph Theory and Network Analysis Group\\Department of Computer Science\\ Technical University of Kaiserslautern\\Gottlieb-Daimler-Str. 48-668\\67663 Kaiserslautern, Germany\\
Telephone: +49 631 205 3340\\
Email: tavassoli@cs.uni-kl.de}
\and
\IEEEauthorblockN{Katharina Anna Zweig}
\IEEEauthorblockA{Graph Theory and Network Analysis Group\\Department of Computer Science\\ Technical University of Kaiserslautern\\Gottlieb-Daimler-Str. 48-670\\67663 Kaiserslautern, Germany\\
Telephone: +49 631 205 3346\\
Email: zweig@cs.uni-kl.de}}
\maketitle
\begin{abstract}
Chat-log data that contains information about sender and receiver of the statements sent around in the chat can be readily turned into a directed temporal multi-network representation. In the resulting network, the activity of a chat member can, for example, be operationalized as his degree (number of distinct interaction partners) or his strength (total number of interactions). However, the data itself contains more information that is not readily representable in the network, e.g., the total number of words used by a member or the reaction time to what the members said. As degree and strength, these values can be seen as a way to operationalize the idea of {\it activity} of a chat-log member. This paper deals with the question of how the overall activity of a member can be assessed, given multiple and probably opposing criteria by using a fuzzy operator. We then present a new way of visualizing the results and show how to apply it to the network representation of chat-log data. Finally, we discuss how this approach can be used to deal with other conflicting situations, like the different rankings produced by different centrality indices.
\end{abstract}
\IEEEpeerreviewmaketitle
\section{Introduction}\label{sec1:Introduction}
In network analysis, there are often multiple ways of measuring a structural property: centrality indices are well known for focusing on very different aspects of the {\it prominence} of a node's position in a graph, weighted networks, temporal networks, or multi-networks further add to this amazing set of measures~\cite{Barrat2004,Borgatti2005,Holme2012,Koschuetzki2005}. For example, for weighted networks, one can now either measure the {\it degree}, the number of interaction partners, or the {\it strength} of a node, i.e., the sum of the weight of incident edges. Since both measures might result in opposing rankings of the nodes, Opsahl et al. argued that there might be situations in which it is necessary to scale between the {\it degree} and the {\it strength} of a node. They proposed a new formula in which a parameter is used to achieve this scaling idea~\cite{Opsahl2010}:
 \begin{equation}
  \label{eq:eq4}
 deg^{\alpha}(i) = deg(i) \times \left(\frac{s(i)}{deg(i)}\right)^\alpha
 \end{equation}
where $deg(i)$ is the degree of node $i$, $s(i)$ is its strength, and $\alpha$ is the scaling parameter: if it is set to $0$, the centrality of $i$ is determined by $deg(i)$ alone. If $\alpha$ is set to $1$, it is determined by $s(i)$ alone. 
However, Opsahl's measure cannot be generalized to more than two opposing rankings. Therefore, in this paper, we use a well-known aggregation operator from fuzzy logic, to deal with all situations in network analysis, where some features of nodes in a network can be measured by multiple methods, whose result is conflicting in assigning a rank to the nodes. The concrete data used in this study is a network deduced from chat-logs of an online group psychotherapy~\cite{Tavassoli2014}. This data contains time stamps for each statement submitted and the sentences themselves which cannot easily be represented and analyzed in a graph; we thus propose to turn this data into features of the nodes and aggregate their values with those of the network analytic measure to enable an automatic, explorative analysis. The following section motivates why it can be necessary to take into account more than only the degree and the strength as features measuring the activity of a person in the concrete example of a chat-log data.
\section{Analyzing the activity of a person using a fuzzy aggregation operator}
\begin{figure}[!t]
\centering
\includegraphics[width=8.5 cm]{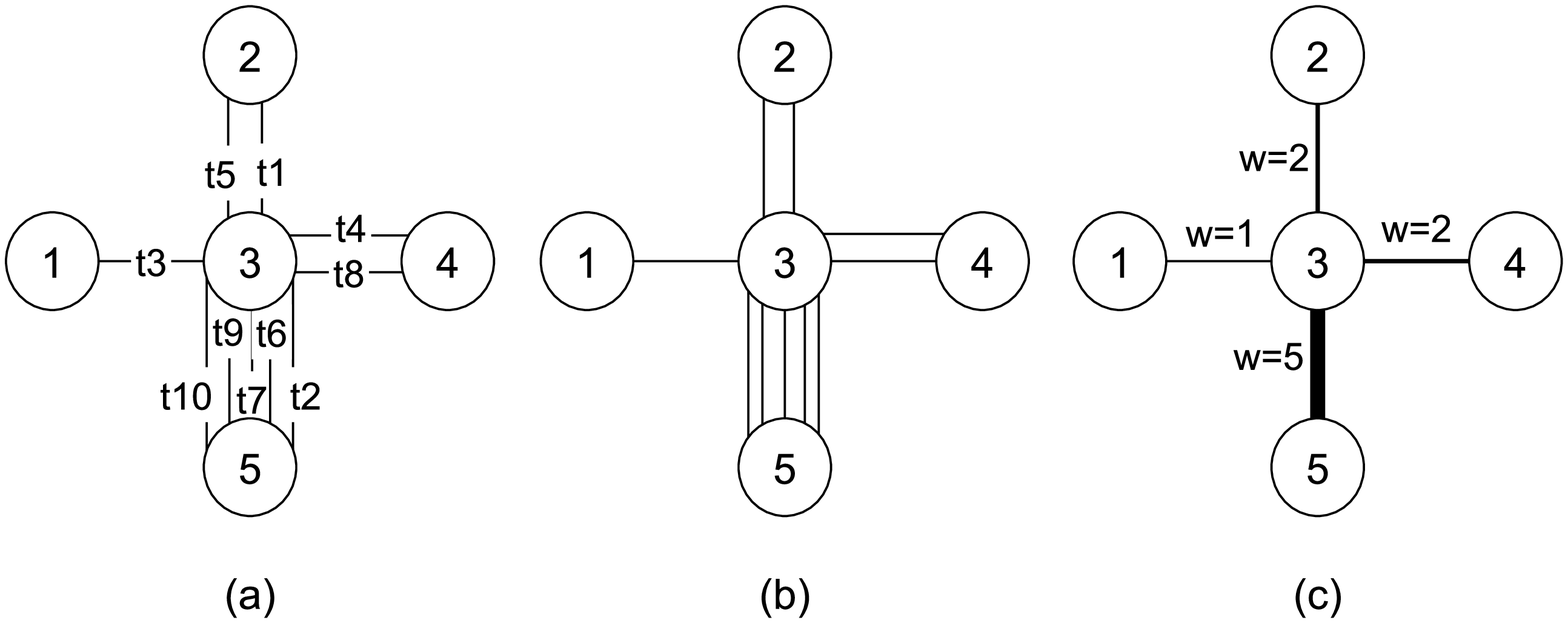}
\caption{\label{fig:figrep} a) a temporal graph whose edges are time-stamped. b) the aggregated form of the graph (a) after ignoring the time sequence of the edges. c) the weighted graph of the graph (b) after summing up the number of edges between a pair of nodes.}
\end{figure}

Recent studies show that the majority of mental disorders can be partly recovered in online group psychotherapy sessions~\cite{Moessner2012}. 
Thus, constructing a social network deduced from large amounts of digitally archived interactions allows for a much more extensive exploration~\cite{Tavassoli2014}. Due to the enormous scale of  available data of more than $2,000$ sessions of at least $60$ minutes, new, explorative measures have to be developed to find interesting patterns that indicate productive sessions and helpful therapists. Since the data also contains more information than just the fact that an interaction happened (i.e., a sentence from user $A$ directed to user $B$), namely the number of words or the exact timing of when the sentence was submitted, measures should be able to use this information as well.
The classic way to deal with temporal interaction data is to aggregate it; in the case of chat-logs of online group therapy, a natural aggregation unit is a single session.
A network representation can be chosen such that each participant is represented by a node and an interaction is represented by an edge. Of course, in a chat, there are in general multiple interactions between the same persons. This can be represented by multi-edges or by edges weighted with the number of interactions (s. Fig.~\ref{fig:figrep}). Note that both kinds of representations should be directed as $A$ might direct more sentence to $B$ then vice versa. The weight could also be more fine-grained by summing up, e.g., the number of words contained in all interactions directed from $A$ to $B$. Based on such a representation, the above mentioned scaling between {\it degree} and {\it strength}, i.e., the total weight, can be measured using Eq.~\ref{eq:eq4}.
However, it seems to be reasonable to also explore measurable features of a single person in the chat, i.e., values associated to the node that might influence our concept of centrality in chat groups. 
In general, the responsibility of the therapist generally demands higher activity compared to the other members~\cite{Yalom2005}. For example, the therapist might need to spend more time in communicating with each other or address more of the other members than a normal member. Sometimes, however, the therapist should talk less than other members and just manage the flow of the chat:  because of the informal and basically unstructured nature of a chat, some members strive for dominating it and others isolate themselves. 
This might lead the therapist to have many interactions with the same persons (low degree, high strength) to either activate or moderate them. 
Without question, in most cases a reasonable measure should identify the therapist(s) as the most central person. 
In this paper, we aim at showing how additional features of a node can be analyzed together with network analytic based values by using a fuzzy aggregation operator, to explore the range of possible rankings that scale between the structural position of a node and its features.
The two important features used in Eq.~\ref{eq:eq4} are also included in the set of the features associated to the nodes in our method. 
One (very simple) way to operationalize the activity of a member is to count the number of words a user submits as one feature. Another way 
is to sum up the reaction time of the other members to his or her statements. 
We operationalize this idea by measuring the {\it total reaction time}, i.e., the total time until the next member starts sending a statement after $i$ submitted a sentence (disregarding the $10$ first and last sentences of a session which only contain hello and farewell statements). 
For illustrating the usage and usefulness of a fuzzy logic operator we thus assign to each node the following four features:
   \begin{itemize}
    \item $a_1(i):$ the number of communication partners ({\it in- plus out-degree}).
		\item $a_2(i):$ the total number of statements the member $i$ sent ({\it out-strength}).
    \item $a_3(i):$ the {\it total number of words} the member $i$ submitted.
    \item $a_4(i):$ the {\it overall reaction time}.
   \end{itemize}	
Note that the specific characteristic of the features is not our main point. In the experimental result, we show that they are just chosen to present some concrete examples on which the benefit of using a fuzzy operator is demonstrated, as described in the following.
\subsection{Ordered Weighted Averaging (OWA)}
The idea of fuzzy operators is to take a set of values and to aggregate them into a single value. 
The aggregation operators are generally classified as either {\it conjunctive} or {\it disjunctive}, depending on whether they combine the values by a logical {\bf and} or an {\bf or} operator respectively. A number of studies show that these operators are very practical in Multi-criteria decision making where a solution for a problem needs to satisfy at least one, all, some, or most of the criteria~\cite{Yager1988,Bellman1970,Zadeh1983}.  
These linguistic terms help to represent what we expect to get from the corresponding aggregation~\cite{Zadeh1983}.
For some of the problems, it does not matter {\bf which} of the criteria are actually satisfied as long as enough of them are satisfied. We will assume, that the degree to which some criteria $j$ is satisfied can be expressed by a real positive number between $0$ and $1$, and that $1$ means full satisfaction and $0$ means no satisfaction. If we require all of the criteria to be fully satisfied, this can be stated as that the least satisfaction of any criterion must be $1$. If we require that at least one criterion must be fully satisfied, this can be stated as that the largest satisfaction of any criterion must be $1$. Thus, the idea of ordered weighted aggregation operator was to order the degree of satisfaction of all criteria and multiply it with a weight vector. Let $A$ be the set of satisfaction values for all criteria, let $A'$ be descending version of it, and $W=[w_1,w_2, ..., w_{m}]$ (where $m=|A|$) be a weight vector with $\sum_i w_i =1$. Then the weight vector $[1,0,0,...0]$ returns the maximal satisfaction value. 
If the weight vector is $[0,0, ..., 1]$, the minimal satisfaction value is returned.
The maximum operator is equated with the crisp existential quantifier (``there exists one criterion that is satisfied'') and with the $OR$ operator; analogously, the minimum operator is equated with the crisp universal quantifier (``all criteria are satisfied'') and with the $AND$ operator~\cite{Yager1988}. 
Therefore, the operator ${OWA}$ is defined as a mapping function $I^{m}\rightarrow I$ (where $I=[0,1]$) as follows:
\begin{equation}
\label{eq:eq2}
{OWA}(a_1,a_2,\cdots , a_m) = \sum_{j=1}^{m} w_j b_j 
\end{equation}
where $b_j$ is the $jth$ largest value of $a_m$.
Of course, now the general idea of ordering the values and weighting can be taken further: for example, setting the weight vector to $[1/m, 1/m, ..., 1/m]$ returns the regular average of the satisfaction. 
The {\it orness} of a weight vector like this is $0.5$, the orness of the $[1,0,0,...,0]$ vector is $1$ and the orness of the $[0,0,0,...,1]$ vector is $0$ 
($andness=1-orness$ as defined in~\cite{Yager1994}). 
Using a weighting function, we can express the concept of a wide range of linguistic quantifiers available between two absolute quantifiers of "there exists" and " for all"~\cite{Zadeh1983}. In addition, any relative quantifier such as must, few, or many, can be represented using a fuzzy subset where the quantifier specifies to which degree a given proportion of criteria satisfies the concept of the chosen linguistic term~\cite{Yager1994,Zadeh1983}. For example, a weight vector $W=[w_1,w_2,...,w_m]$ can be defined using a function $Q(r)=r^{\beta}$ with a parameter $\beta\geq 0$ (an increasing monotone quantifier) in the following way:
\begin{equation}
w_i = Q\left(\frac{i}{m}\right)-Q\left(\frac{i-1}{m}\right)
\end{equation}
It can be easily seen that this vector sums to $1$, since all terms of the form $Q\left(\frac{i}m\right)$ cancel each other for $ 1 \leq i \leq m-1$, and $Q(m/m)=Q(1)=1^\beta=1$ and $Q(0/m)=0$. The {\it orness} of such a vector is: \[orness(Q_\beta)= \int_{0}^{1} r^{\beta} dr=\frac{1}{\beta + 1}\] Thus, if $\beta > 1$ then $orness < 0.5$ and if $\beta < 1$ then $orness > 0.5$~\cite{Yager1994}. 
The value of the $\beta$ varies the property of compensation among $(a_1,a_2,...a_m)$ from full compensation (high-orness) to no compensation (high-andness). This means that a higher degree of satisfaction of one of the features can compensate for a lower degree of satisfaction of another features. Note that it is crucial to make sure that there is a weak order relation on the set of features for the nodes. Otherwise, the features place in the same positions after sorting in Eq.~\ref{eq:eq2} and get the same importance corresponding to the weight vector; this might lead a feature to have a high importance in all aggregations and cause that it matters {\bf which} of the features is satisfied for measuring the activity of a person. 
\section{Experimental Result}
\begin{figure*}[!thp]
  \centering
  \includegraphics[width=7.2 in]{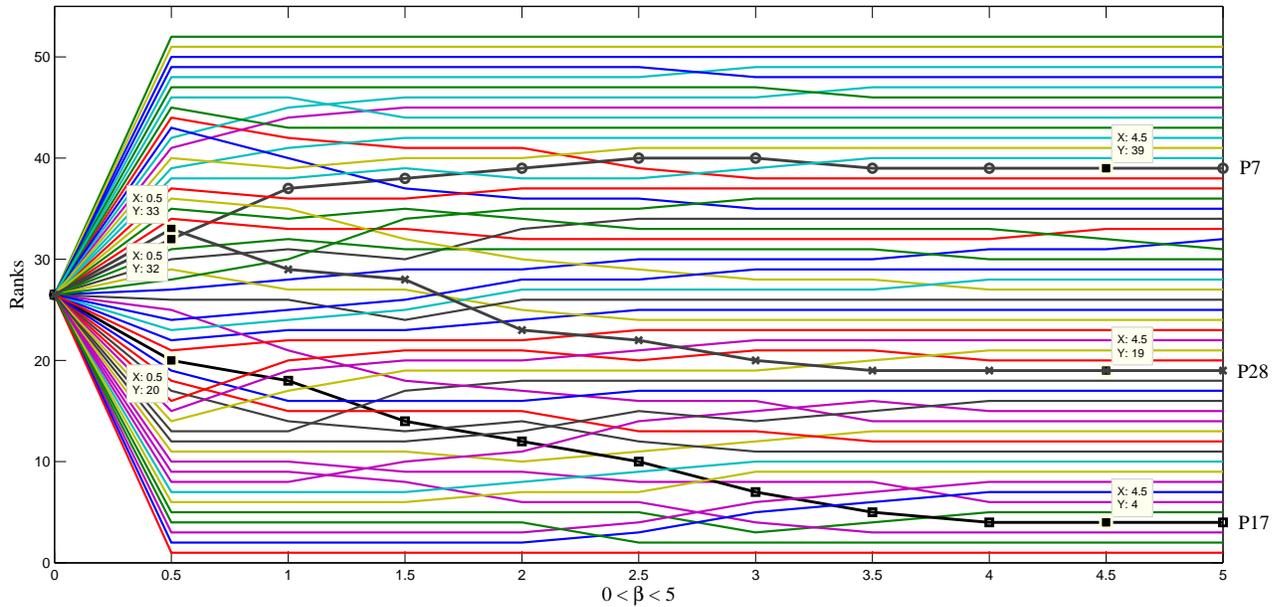}
  \caption{\label{fig1:session25} {This figure shows how the ranks of the nodes from the highest $52$ to the lowest $1$ change over the different values of $\beta$. The possible values of $\beta$ vary in the interval $[0,5]$ in this example and result in the andness $[0,0.8333]$ which indicates how the features can compensate each other's value from full compensation (high-orness) to no compensation (high-andness). Three chosen samples are marked who have very strong differences in their ranks in the different values of $\beta$. These nodes are highlighted in Table.~\ref{tab1:Activity} to see their ranks using only the normalized \textit{degree}.}}
  \end{figure*}

To show the explorative power of the OWA operator on the different activity features, we used a network including $52$ nodes and $29590$ multi-edges labeled by time stamps from one of the group-chat sessions.
In order to put the features on the same scale, the features are normalized by their respective maximum and minimum values into $[0,1]$. 
 We computed the outcome for different values of the {\it orness}-parameter $\beta$ and ranked the nodes by it. Remember that values close to $0$ indicate highest orness, $\beta=1$ reproduces the average of the four activity features and $\beta=\infty$ indicates an orness of $0$ (an andness of $1$). Fig.~\ref{fig1:session25} shows the result: the most obvious and expected result is that the therapist is ranked highest for all values of $\beta$, i.e., independent of whether the operator favors the highest activity value or the lowest or any mixture of it, the therapist is the most active. The figure also shows that most participants' rank is quite stable, independent of the {\it orness} of the OWA operator. However, three nodes show a very strong difference in their ranking on the extreme scales of $\beta$: the first one is $P7$. He communicates with only $9$ people out of the $52$ members of the chat-log. If he was only ranked by the normalized degree, this would be a medium value (rank 28 out of 52) as listed in Table.~\ref{tab1:Activity}. However, his activity with those $9$ persons, measured in the number of words typed, is quite high as his activity with respect to the number of statements and the total reaction time. In all of these other three values, he is ranked among the first quarter of ranks. Since he is never on the top rank of any of the measures, a high-orness makes him a {\it medium active} person. Decreasing the orness and increasing the andness, however, shows that he has a very stable medium to high rank in {\bf all} of the criteria. $P17$'s case is even more interesting: he has more distinct communication partners, namely $10$. This gives him a higher rank in the high-orness (the low-andness) than $P7$. However, all his other activity features are extremely close to $0$, such that the higher the andness of the operator, the lower his ranking. Looking into the session, it turns out that $P17$ is not actually a patient, but a psychologist who visited the chat-log and was a mere spectator. The last case, $P28$, has a higher rank than $P7$ using only the degree of the nodes for ranking as listed in Table.~\ref{tab1:Activity}. The same situation he has in the high-orness (the low-andness) as depicted in Fig.~\ref{fig1:session25} but once the degree of compensation is decreased by increasing the value of $~\beta$, his rank is gradually dropped from $33$ to $19$. Going through the chat-log, it shows that he has communicated with more distinct members than $P7$ using very short, say, single word statements consecutively. Thus, considering a high-andness on the set of his activity features decreases his rank very fast as he needs to have most of them or all of them satisfied. We did not know this in advance and randomly pick this session out of the $2,000$ available sessions. 
\section{Conclusion and future work}
Network analysis often has to deal with a multitude of information known about a node: in this paper, we concentrated on some situations in which a node was assigned both structural features classically known from network analysis (degree and strength) together with data not directly present in the network (number of words, total reaction time). While the four features all focus on one possible operationalization of {\it activity}, they also create opposing ranks. The exploratory analysis with the OWA operator from fuzzy logic helps to understand how strong the ranks oppose, and thus give more insight then, e.g., a pairwise correlation of the values. Our examples identified some cases whose ranks got extremely changed by applying more features to the analysis of their activity. Looking into the chat-log data set, we found the main reasons which could not be recognized in the ranking using only two features the \textit{degree} or the \textit{strength} in such analysis. 
 
 It is immediately clear that the proposed method can also help to deal with other conflicting information around nodes' position in the network: (1) Consider as features the degree of a node in various time intervals. E.g., each session could be cut into intervals of $10$ minutes, and the degree is measured for each of these intervals. Different values of $\beta$ then identify whether someone is active for only some interval or for almost all intervals. (2) Consider as features the classic centrality values, e.g., degree, closeness, and betweenness. Here, different $\beta$ values scale between the node who is most active in one of them and the one who has the highest value in all of them. 
\begin{table}[!bp]
\centering
\caption{Top $20$ nodes are listed from the highest rank $52$ to the lowest $20$ with respect to the result of $a_1(i)$ which is the normalized \textit{degree} of node $i$. Three nodes those which have some particular changes in their ranks in Fig.~\ref{fig1:session25} are highlighted here.} 
 \label{tab1:Activity}
\begin{tabular}{c|ccccc}
\cline{2-6}
\multicolumn{1}{c|}{Rank} & \multicolumn{1}{c|}{$i$} & \multicolumn{1}{c|}{$a_1(i)$} & \multicolumn{1}{c|}{$a_2(i)$} & \multicolumn{1}{c|}{$a_3(i)$} & \multicolumn{1}{c|}{$a_4(i)$} \\\cline{1-6} 
52                        & Therapist                     & 1                              & 1                             & 1                               & 1                               \\
51                        & P6                     & 0.588                          & 0.423                         & 0.196                           & 0.407                           \\
50                        & P1                     & 0.51                           & 0.276                         & 0.224                           & 0.257                           \\
49                        & P36                    & 0.49                           & 0.265                         & 0.168                           & 0.249                           \\
48                        & P37                    & 0.451                          & 0.217                         & 0.145                           & 0.2                             \\
47                        & P11                    & 0.431                          & 0.229                         & 0.204                           & 0.205                           \\
46                        & P22                    & 0.431                          & 0.044                         & 0.064                           & 0.037                           \\
45                        & P3                     & 0.392                          & 0.116                         & 0.042                           & 0.109                           \\
44                        & P30                    & 0.392                          & 0.104                         & 0.136                           & 0.105                           \\
43                        & P39                    & 0.373                          & 0.161                         & 0.145                           & 0.151                           \\
42                        & P27                    & 0.333                          & 0.033                         & 0.02                            & 0.032                           \\
41                        & P24                    & 0.314                          & 0.061                         & 0.048                           & 0.059                           \\
\rowcolor[HTML]{C0C0C0} 
40                        & P28                    & 0.314                          & 0.01                          & 0.008                           & 0.011                           \\
39                        & P41                    & 0.314                          & 0.092                         & 0.073                           & 0.083                           \\
38                        & P5                     & 0.294                          & 0.183                         & 0.152                           & 0.173                           \\
37                        & P10                    & 0.294                          & 0.031                         & 0.044                           & 0.031                           \\
36                        & P18                    & 0.294                          & 0.182                         & 0.167                           & 0.168                           \\
35                        & P23                    & 0.294                          & 0.043                         & 0.027                           & 0.043                           \\
34                        & P25                    & 0.294                          & 0.091                         & 0.066                           & 0.09                            \\
33                        & P4                     & 0.275                          & 0.102                         & 0.139                           & 0.096                           \\
32                        & P16                    & 0.255                          & 0.033                         & 0.056                           & 0.028                           \\
31                        & P14                    & 0.235                          & 0.043                         & 0.037                           & 0.039                           \\
30                        & P20                    & 0.235                          & 0.024                         & 0.014                           & 0.02                            \\
29                        & P15                    & 0.216                          & 0.034                         & 0.03                            & 0.035                           \\
\rowcolor[HTML]{C0C0C0} 
28                        & P7                     & 0.196                          & 0.127                         & 0.056                           & 0.104                           \\
27                        & P33                    & 0.196                          & 0.003                         & 0.003                           & 0.003                           \\
26                        & P51                    & 0.196                          & 0.063                         & 0.04                            & 0.053                           \\
\rowcolor[HTML]{C0C0C0} 
25                        & P17                    & 0.176                          & 0                             & 0                              & 0                              \\
24                        & P42                    & 0.176                          & 0.027                         & 0.02                            & 0.026                           \\
23                        & P8                     & 0.157                          & 0.022                         & 0.024                           & 0.02                            \\
22                        & P31                    & 0.157                          & 0.003                         & 0.004                           & 0.002                           \\
21                        & P35                    & 0.157                          & 0.002                         & 0.003                           & 0.003                           \\
20                        & P43                    & 0.157                          & 0.004                         & 0.006                           & 0.004                                                   
\end{tabular}
\end{table} 
\bibliographystyle{plainnat}
\bibliography{SudesSecondBib}
\end{document}